\documentclass[11 pt,nofootinbib,notitlepage,eqsecnum]{revtex4-1}
\usepackage[english]{babel}
\usepackage{bbm}
\usepackage{mathtools} 
\usepackage{hyperref}
\usepackage{soul,xcolor}
\setstcolor{red} 
\hypersetup{
    colorlinks,
    linkcolor={red!50!black},
    citecolor={blue!50!black},
    urlcolor={blue!80!black}
}
\addto\extrasenglish{%

}

\showboxdepth=\maxdimen
\showboxbreadth=\maxdimen

\usepackage{subfig}
\usepackage{graphicx,tikz}
\usetikzlibrary{arrows,shapes.geometric,positioning,shapes.misc}
\usepgfmodule{shapes}
\usetikzlibrary{decorations.markings,decorations.text,decorations,patterns,calc}
\usepackage{pgfkeys}

\usepackage{amsmath,amssymb,amsthm,MnSymbol}

\newcommand{\dd}{\mathop{}\!d}

\newcommand{\eqat}[1]{\mathrel{\text{\setbox0\hbox{$=$}\rlap{\hbox to \wd0{\hss\raisebox{-.70\height}{$\scriptscriptstyle #1$}\hss}}\box0}}}

\def\Re{\mathrm{Re}\,}

\def\tr{\mathrm{tr}}
\def\half{\frac{1}{2}}
\def\sgn{\mathrm{sgn}}
\def\eps{\epsilon}

\newcommand{\ket}[1]{|#1\rangle}
\newcommand{\bra}[1]{\langle #1|}
\newcommand{\scal}[1]{\langle #1\rangle}

\def\tr{\mathrm{tr}}

\def\R{\mathbb{R}}

\def\C{\mathbb{C}}
\def\CP{\mathbb{CP}}

\newcommand{\SL}{\mathrm{SL}(2,\C)}

\newcommand{\Spin}{\mathrm{Spin(4)}}

\newcommand{\SU}{\mathrm{SU(2)}}

\newcommand{\EPRL}{\mathrm{EPRL}}

\def\EuclideanEPRL{\cite{epr2007,epr2008,elpr2008}}
\def\asymptoticL{\cite{bdfhp2010}}
\def\asymptoticE{\cite{bdfgh2009}}
\def\jonProp{\cite{engle2013}}
\def\jonCorr{\cite{engle2013d}}
\def\jon{\cite{engle2011a,engle2013d,engle2013a,engle2013}}
\def\brv{\cite{brv2010}}
\def\cohstates{\cite{almmt1996, thiemann2001, thiemann2006a, bmp2009}}
\def\propcit{\cite{bmrs2006, ar2007, ar2008, bmp2009a, bd2012, cev2015}}

\def\beq{\begin{equation}}
\def\eeq{\end{equation}}
\def\bq{\begin{equation*}}
\def\eq{\end{equation*}}

\newcommand{\propdec}{\mathrm{(+)}}
\newcommand{\Sprop}{S^{\propdec}}
\newcommand{\Aprop}{A^{\propdec}_v}
\newcommand{\Wprop}{W^{\propdec}}
\newcommand{\Wcor}{W^{\propdec}_{\mathrm{corr}}}

\begin{document}
\title{Spinfoam cosmology with the proper vertex amplitude}

\author{Ilya Vilensky}
\email{ilya.vilensky@fau.edu}

\affiliation{\mbox{Department of Physics, Florida Atlantic University, Boca Raton, Florida, USA}}

\begin{abstract}
\centerline{\bf Abstract}

The proper vertex amplitude is derived from the EPRL vertex by restricting to a single gravitational sector in order to achieve the correct semi-classical behaviour. We apply the proper vertex to calculate a cosmological transition amplitude that can be viewed as the Hartle-Hawking wavefunction. To perform this calculation we deduce the integral form of the proper vertex and use extended stationary phase methods to estimate the large-volume limit. We show that the resulting amplitude satisfies an operator constraint whose classical analogue is the Hamiltonian constraint of the Friedmann-Robertson-Walker cosmology. We find that the constraint dynamically selects the relevant family of coherent states and demonstrate a similar dynamic selection in standard quantum mechanics. We investigate the effects of dynamical selection on long-range correlations.

\end{abstract}
\maketitle

\section{Introduction}
Spinfoam models provide a path integral description of the dynamics of loop quantum gravity (LQG), a proposed theory of quantum gravity. The most widely studied model is the Engle-Pereira-Rovelli-Livine (EPRL) vertex amplitude {\EuclideanEPRL}. However, it has been pointed out that this model fails to select a single gravitational sector {\jonCorr} which may lead to unphysical contributions in the semi-classical limit from configuration histories that do not satisfy the classical equations of motion. A proposed modification of the vertex amplitude that resolves this issue by introducing a quantum mechanical restriction to a single gravitational sector has been developed under the name of the 'proper' vertex amplitude {\jon}.

One of the most important tasks before any theory of quantum gravity is to provide a description of the universe near the Big Bang singularity, in the regime where classical equations of general relativity break down. Within the LQG framework loop quantum cosmology (LQC) has seen the most development. In this approach one starts with a symmetry-reduced model on the classical level and then implements loop quantisation techniques to obtain a theory of (symmetry-reduced) quantum geometry. Another approach, that we take in this work, is to start with the full theory and apply it to a cosmological model. Given the spinfoam dynamics, quantum transition amplitudes can be calculated, giving rise to spinfoam cosmology. The definition and interpretation of transition amplitudes in a background-independent theory of quantum gravity is subtle: see, for example, a recent work on black hole dynamics \cite{crsv2016}. Bianchi, Rovelli and Vidotto {\brv} studied transition amplitudes defined by the EPRL vertex and demonstrated that there is an approximation leading to the classical Friedmann-Robertson-Walker (FRW) cosmology.

In the current paper we investigate quantum amplitudes using the Euclidean proper vertex. Thus, this work provides another test of the proper vertex which has been previously used (in its Lorentzian guise) to calculate the graviton propagator \cite{cev2015}.  The calculation begins with fixing a graph. We choose the boundary states to be based on the graph with five nodes and ten links which can be viewed as a boundary of the $4$-simplex. This boundary graph truncates the Hilbert space of the theory to a finite number of degrees of freedom. The boundary is seen to be a $3$-dimensional slice of a homogeneous and isotropic universe. Then we pick as the boundary states the coherent states peaked on the intrinsic and extrinsic geometry of a spatial slice of FRW spacetime (such coherent states were also considered in \cite{mmp2011}). These coherent states also encode the quantum fluctuations around the FRW geometry and their dynamics includes some inhomogeneous and anisotropic degrees of freedom.

We work at first order in the vertex expansion. The resulting quantum amplitude can be interpreted as the transition amplitude from a zero three-geometry to a compact three-geometry. Such amplitude has been proposed by Hartle and Hawking {\cite{hh1983}} as a quantum ground state of the universe, termed the Hartle-Hawking wavefunction. We proceed by evaluating the proper vertex amplitude in the coherent state representation. We estimate this amplitude in the large-volume limit. This allows us to use stationary phase methods to obtain an approximation for the Hartle-Hawking wavefunction. 

To make a connection with the classical model we show that the amplitude $\Wprop$ satisfies the operator constraint $\hat{H} \Wprop = 0$. We demonstrate that its classical analogue is the classical Hamiltonian constraint that arises in LQC. The dynamics of the model is found to select a particular family of coherent states. We shed light on this restriction by drawing an analogy with a similar dynamical selection in standard quantum mechanics. Then we modify the ansatz to generate long-range correlations and derive a restriction on parameters characterising these correlations.

The paper is organized as follows. In \autoref{sec:prelim} the definitions of both EPRL and proper vertex amplitudes are reviewed. In \autoref{sec:cosmo} the approximations are presented and the amplitude is evaluated. In \autoref{sec:classlim} we analyse the classical limit and dynamical restrictions on the set of coherent states. In \autoref{sec:corr} we investigate the effects of dynamics on long-range correlations. We close with a summary of the results and a discussion of future work. 

\section{Preliminaries}
\label{sec:prelim}
\subsection{EPRL vertex}
We recall the $\Spin$ EPRL vertex amplitude defined on a given oriented $4$-simplex. The tetrahedra have labels running from $0$ to $4$ which we denote $a$, $b$. The boundary Hilbert space is spanned by $\SU$ generalised spin network states $\Psi$ labelled by spins $j_{ab}$ and vectors $\psi_{ab}$, $\psi_{ba}$ in the corresponding irreducible representation of $\SU$, defined explicitly by $\Psi(\{U_{ab}\})=\prod_{a<b}\langle \psi_{ab}|U_{ab}| \psi_{ba} \rangle $ with $a$, $b$ taking values in the range from $0$ to $4$.

Let $V_j$ denote the representation space for the spin $j$ representation of $\SU$ which will be denoted by $\rho_j(g)$ for $g\in\SU$ (the $j$ subscript will be omitted when it is clear from the context). Let $\hat{L}^i$ denote the generators in each of these representations. Let $\eps: V_j \times V_j \rightarrow \C $ be the invariant bilinear inner product and $\langle \cdot,\cdot \rangle$ the Hermitian inner product on $V_j$. An antilinear structure map $J: V_j \rightarrow V_j$ is then given by $\eps(\psi,\phi)=\langle J\psi, \phi \rangle$. $J$ commutes with the group representation matrices and anticommutes with the generators.

Now let $V_{j^+,j^-} = V_{j^+} \otimes V_{j^-}$ denote the representation space for the spin $(j^+,j^-)$ representation of $\Spin=\SU \times \SU$ and $\rho_{j^+,j^-}(X^+,X^-) := \rho_{j^+}(X^+) \otimes \rho_{j^-}(X^-)$ denote the representation of $(X^+,X^-)\in\Spin$ (again with the subscripts dropped when clear from the context). Define the bilinear form $\eps: V_{j^+,j^-} \times V_{j^+,j^-} \rightarrow \C$ by $\eps(\psi^+\otimes\psi^-,\phi^+\otimes\phi^-) := \eps(\psi^+,\phi^+)\eps(\psi^-,\phi^-)$ and the antilinear map $J: V_{j^+,j^-} \rightarrow V_{j^+,j^-}$ by $J(\psi^+\otimes\psi^-) = (J\psi^+)\otimes(J\psi^-)$. Then $\eps(\Psi,\Phi)=\langle J\Psi,\Phi \rangle$. Finally, let $Y^{j^+,j^-}_j: V_j \rightarrow V_{j^+,j^-} $ denote the Clebsch-Gordan intertwining map.

A group element $G_a = (X^+_a,X^-_a)$ is assigned to each tetrahedron $a$ in the boundary of the $4$-simplex. Define $G_{ab} := (G_a)^{-1}G_b$. This group element can be interpreted for each pair of tetrahedra $a$, $b$ as the parallel transport map $G_{ab}= (X^+_{ab},X^-_{ab})$ from the frame of tetrahedron $b$ to the frame of tetrahedron $a$.  

The imposition of the linear simplicity constraint fixes
\bq 
j^{\pm}_{ab} = \frac{|1\pm\gamma|}{2} j_{ab}
\eq
where $\gamma$ is the Barbero-Immirzi parameter. Then the EPRL vertex amplitude for a given LQG boundary state $\Psi_{\{j_{ab},\psi_{ab}\}}$ is
\beq
A_v(\Psi_{\{j_{ab},\psi_{ab}\}}) = \int_{\Spin^5} \prod_a \dd G_a \prod_{a<b}\eps(Y^{j^+_{ab},j^-_{ab}}_{j_{ab}}\psi_{ab},\rho(G_{ab})Y^{j^+_{ab},j^-_{ab}}_{j_{ab}}\psi_{ba}).
\eeq

In this paper we will use the coherent state formulation of the vertex amplitude where instead of the vectors $\psi_{ab}$, $\psi_{ba}$ the boundary spin-network states are labelled by the Perelomov coherent states \cite{perelomovGCS} $C^{j_{ab}}_{\xi_{ab}}$, $C^{j_{ab}}_{\xi_{ba}}$ associated with unit spinors $\xi_{ab}$, $\xi_{ba}$. We also define a unit $3$-vector $n_{\xi}$, corresponding to a $2$-spinor $\xi$, by
\bq
n_{\xi} := \frac{\langle \xi| \sigma | \xi \rangle}{\langle \xi | \xi \rangle} .
\eq
For any normalised spinor $\xi$ take 
\bq 
g(\xi)=\begin{pmatrix}\xi_0&-\overline{\xi_1}\\\xi_1&\overline{\xi_0}\end{pmatrix} \in \SU .
\eq
Then the coherent state $C^j_{\xi}$ is given by
\bq 
C^j_{\xi} := g(\xi)\ket{j,j}, 
\eq
that is, the highest weight eigenstate of $n_{\xi} \cdot \hat{L}$ in the spin $j$ representation. Therefore, the EPRL vertex amplitude on the coherent states is
\beq 
A_v(\{j_{ab}, C^{j_{ab}}_{\xi_{ab}} \}) = \int_{\Spin^5} \prod_a \dd G_a \prod_{a<b}\eps(Y^{j^+_{ab},j^-_{ab}}_{j_{ab}}C^{j_{ab}}_{\xi_{ab}},\rho(G_{ab})Y^{j^+_{ab},j^-_{ab}}_{j_{ab}}C^{j_{ab}}_{\xi_{ba}}).
\eeq

\subsection{Proper vertex}
When the boundary data defines a non-degenerate $4$-simplex geometry, Barrett \textit{et al.} show that the EPRL vertex amplitude contains four terms in the semi-classical limit {\asymptoticE}. In a series of papers {\jon} Engle introduced the proper vertex amplitude and showed that its semi-classical limit comprises only one term with the Regge action appearing with the positive sign. This amplitude is defined by
\beq 
\Aprop = \int_{\Spin^5} \prod_a \dd G_a \prod_{a<b} \eps(Y^{j^+_{ab},j^-_{ab}}_{j_{ab}}\psi_{ab},\rho(G_{ab})Y^{j^+_{ab},j^-_{ab}}_{j_{ab}}\Pi_{ba}(\{G_{a'b'}\})\psi_{ba})
\eeq
where $\Pi_{ba}(\{G_{a'b'}\})$ is a projection operator acting in the spin $j_{ab}$ representation of $\SU$, given by
\beq 
\Pi_{ba}(\{G_{a'b'}\}) := \Pi_{(0,\infty)} \left(\beta_{ab}(\{G_{a'b'}\})\tr(\sigma_i X^-_{ab}X^+_{ba})\hat{L}^i\right).
\eeq
Here $ \Pi_{(0,\infty)}(\hat{O})$ denotes the spectral projector onto the positive part of the spectrum of the operator $\hat{O}$,  
\bq
\beta_{ab}(\{G_{a'b'}\}) = -\sgn[\eps_{ijk}n^i_{ac} n^j_{ad} n^k_{ae} \eps_{lmn} n^l_{bc} n^m_{bd} n^n_{be}],
\eq
with $\{c,d,e\} = \{0,\dotsc,4\}\backslash\{a,b\}$, and
\bq
n^i_{ab} = \tr(\sigma^i X^-_{ab}X^+_{ba}).
\eq
The amplitude can be written using coherent states on the boundary as
\beq
\label{eq:Aprop}
 \Aprop(\{j_{ab}, C^{j_{ab}}_{\xi_{ab}} \}) = \int_{\Spin^5} \prod_a \dd G_a \prod_{a<b}\eps(Y^{j^+_{ab},j^-_{ab}}_{j_{ab}}C^{j_{ab}}_{\xi_{ab}},\rho(G_{ab})Y^{j^+_{ab},j^-_{ab}}_{j_{ab}}\Pi_{ba}(\{G_{a'b'}\})C^{j_{ab}}_{\xi_{ba}}). 
\eeq

\section{Cosmological set-up}
\label{sec:cosmo}
\subsection{Choice of a graph and the boundary states}
\begin{figure}
	\includegraphics[scale=1]{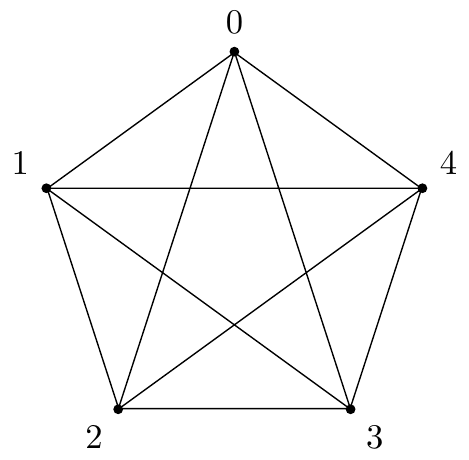}
	\caption{\label{fig:4simplex}The boundary graph $\Gamma_5$.}
\end{figure}

Spinfoam vertex amplitudes give path-integral transition amplitudes for fixed boundary states. To perform this calculation we choose a graph thereby truncating the boundary Hilbert space. Specifically, we choose a graph $\Gamma_5$ formed by five nodes connected with ten links (see \autoref{fig:4simplex}).
This graph can be seen as the boundary of a $4$-simplex. It can be endowed with a geometrical interpretation as follows. Consider a compact connected $3$-manifold $M$ with the topology of a $3$-sphere. Then the boundary of a $4$-simplex can be viewed as a triangulation of $M$. The $3$-manifold $M$ represents a spatial slice of a homogeneous and isotropic universe.

The next step involves picking the boundary states. These LQG states should be peaked on both extrinsic and intrinsic geometry of the $3$-manifold and, therefore, are superpositions of spin networks. Such states are known in the literature {\cohstates} and given explicitly by
\beq
\label{eq:hkcs}
\Psi_{H_l}(U_l) = \int_{\SU^N} \dd g_n \prod_l K_t(g^{-1}_{s(l)} U_l g_{t(l)} H^{-1}_l)
\eeq
where $K_t$ is the heat kernel function of the form
\beq
K_t(g) = \sum_j d_j e^{-t\hbar j(j+1)} \tr(D^j(g)),
\eeq
with $d_j = 2j+1$ and $D^j(g)$ the Wigner matrix in the spin-$j$ representation of $\SU$. Here we insert $\hbar$ into the heat kernel function to ensure that absolute (and relative) uncertainties in both area and its conjugate variable have the same dependence on the Planck constant. The labels $H_l$ appearing above are elements of $\SL$ and can be written as \cite{fs2010}
\beq
H_l = n_{s(l)} e^{-i(\xi_l + i\eta_l)(\sigma_3/2)} n^{-1}_{t(l)},
\eeq
with $n_{s(l)}$, $n_{t(l)}$ elements of $\SU$. In these definitions $s(l)$, $t(l)$ denote, respectively, the source node and the target node of the link $l$ of the boundary graph.

As shown in {\brv}, homogeneity and isotropy lead to $n_{s(l)}=n_{t(l)}=n_l$ and $\xi_l$, $\eta_l$ being independent of $l$. Bianchi, Rovelli and Vidotto derive the relationship between the LQG conjugate variables $A$, $E$ and the boundary state labels $\xi$, $\eta$. Specifically, after identification of the $3$-manifold $M$ with the group manifold of $\SU$, the Killing form $\mathring{q}_{ab}$ can be viewed as the fiducial metric and left-invariant vector fields on $\SU$ as the fiducial triads $\mathring{e}$ (with $\mathring{\omega}$ the corresponding co-triads). Then,\footnote{In {\brv} the authors use slightly different definitions of $(c,p)$, here we use the definitions standard in LQC. The extra factors can be absorbed into constants $\alpha$, $\beta$.}
\beq  
A = c \mathring{V}^{-1/3} \mathring{\omega} \quad\quad\quad E = p \mathring{V}^{-2/3} \sqrt{\mathring{q}} \mathring{e},
\eeq
with $\mathring{V}$ the fiducial volume and $\mathring{q}$ the determinant of the fiducial metric, and {\brv}
\beq
\label{eq:xieta}
\xi_l = \xi = \alpha c \quad\quad\quad \eta_l = \eta = \beta p ,
\eeq
with $\alpha$, $\beta$ certain constants. Thus, the homogeneous and isotropic boundary states can be labelled equivalently by $\xi$, $\eta$ or $c$, $p$.

Introducing the holomorphic variable $z$
\beq 
\label{eq:zdef}
z = \xi + i \eta ,
\eeq
we can write
\beq
\label{eq:Hl}
H_l = n_l e^{-iz\sigma_3/2} n^{-1}_l .
\eeq

\subsection{Hartle-Hawking wavefunction}
\begin{figure}[!ht]
	\includegraphics[scale=1]{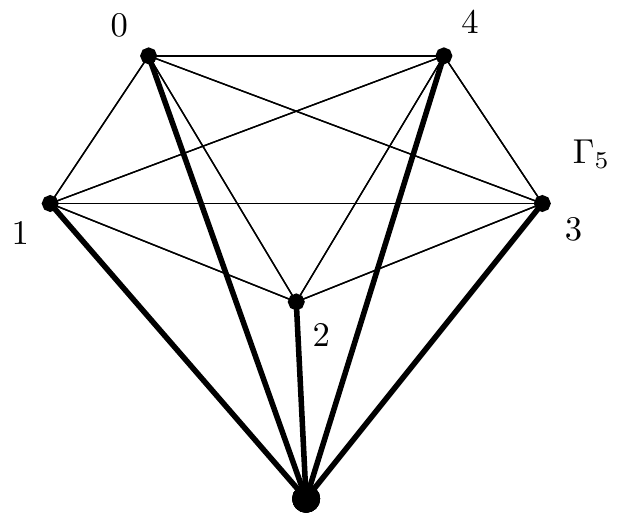}
	\caption{\label{fig:1v4simplex}Spinfoam with a single vertex.}
\end{figure}

In \cite{hh1983} Hartle and Hawking proposed that the wavefunction for a three-geometry is given by the path integral over all compact four-geometries with this three-geometry as a boundary. Spinfoam dynamics of LQG boundary states allows us to implement this proposal. Specifically, we consider an amplitude given by the spinfoam formed from a single vertex bounded by five edges (see \autoref{fig:1v4simplex}).

Such an amplitude can be written as a holomorphic function of $z$ 
\beq
\label{eq:wz}
\Wprop(z) = \langle \Wprop | \Psi_{H_{ab}(z)} \rangle
\eeq
where $\Wprop$ indicates the use of the proper vertex amplitude $\Aprop$. The links of the boundary graph are now labelled by pairs of indices $(ab)$, with $a,b$ denoting the nodes of the graph corresponding to the tetrahedra as described in \autoref{sec:prelim}. We note here that this particular choice of the spinfoam is motivated partly by the fact that the proper vertex amplitude has so far only been defined for $2$-complexes dual to triangulations made up of $4$-simplices.

The amplitude $\Wprop(z)$ can be viewed as a transition amplitude from a zero three-geometry (a single point) to the three-geometry specified by $z$ (with a finite scale factor and extrinsic curvature). It can be rewritten as
\beq
\label{eq:wzint}
\Wprop(z) = \int_{\SU^{10}} \dd U_{ab} \Wprop(U_{ab}) \Psi_{H_{ab}(z)}(U_{ab}).
\eeq

\subsection{Large volume limit}
\label{ssec:calc}
We will calculate the amplitude \eqref{eq:wz} in the large volume limit. This limit is obtained by taking $p$ large or equivalently considering $\eta\gg1$. Using \eqref{eq:Hl}, we write 
\beq
D^j(H_l) = D^j(n_l) D^j(e^{-iz\sigma_3/2}) D^j(n^{-1}_l).
\eeq
In the large $\eta$ limit, we then have \cite{bmp2009}
\beq
D^j\left(e^{-iz\sigma_3/2}\right) \approx e^{-izj}\ket{j,j}\bra{j,j}.
\eeq
Therefore, rewriting \eqref{eq:hkcs}, we get
\beq
\Psi_{H_l}(U_l) \approx \sum_{j_l} \left(\prod_l d_{j_l} e^{-t\hbar j_l(j_l+1) - izj_l} \right) \int_{\SU^N} \dd h_n \prod_l \scal{C^{j_l}_{\xi_l}| h^{-1}_{s(l)}  U_l h_{t(l)} | C^{j_l}_{\xi_l}} .
\eeq
Here the unit spinors $\xi_l$ are chosen to satisfy $n_{\xi_l} = n_l$. 

Plugging this expression into \eqref{eq:wzint}, so that the links of the graph are now labelled by $(ab)$ instead of $l$, performing integrals over $U_{ab}$ and using the invariance of $\Spin$ measure, we obtain
\beq
\label{eq:wpropgauss}
\Wprop(z) =  \sum_{j_{ab}} \left(\prod_{a<b} d_{j_{ab}} e^{-t\hbar j_{ab}(j_{ab}+1) - izj_{ab}}\right) \Aprop\left(\left\lbrace j_{ab}, C^{j_{ab}}_{\xi_{ab}} \right\rbrace\right) .
\eeq
In the limit $\eta \gg 1$ the Gaussian form of the prefactor picks out large values of $j_{ab}$. Therefore, we can evaluate the amplitude factor in the large spin limit. Here the large spin limit is taken by setting all ten spins equal $j_{ab} = j$ and scaling $j \rightarrow \lambda j = j_0$. We will use the extended stationary phase theorem to obtain the asymptotic limit for large $\lambda$. 

To apply stationary phase methods, we first rewrite the amplitude in an exponentiated form. Inserting the completeness relation for coherent states $C^j_{\eta}$ into \eqref{eq:Aprop}, we obtain:
\begin{align}
\Aprop\left(\left\lbrace j_{ab}, C^{j_{ab}}_{\xi_{ab}} \right\rbrace\right) = \int_{\Spin^5} &\prod_a \dd G_a \int_{\CP^{10}}  \prod_{a<b}  \dd \mu_{\eta_{ba}} \nonumber \\ 
&\eps(Y^{j^+_{ab},j^-_{ab}}_{j_{ab}}C^{j_{ab}}_{\xi_{ab}},\rho(G_{ab})Y^{j^+_{ab},j^-_{ab}}_{j_{ab}}C^{j_{ab}}_{\eta_{ba}})\scal{C^{j_{ab}}_{\eta_{ba}}|\Pi_{ba}(\{G_{a'b'}\})C^{j_{ab}}_{\xi_{ba}}} 
\end{align}
where $\eta_{ba}$ are unit spinors (that is, for each of the ten pairs $(ba)$ we have $\scal{\eta_{ba}|\eta_{ba}}=1$ ) and $\dd \mu_{\eta_{ba}} = \frac{d_{j_{ab}}}{\pi} \Omega_{\eta_{ba}}$ with $\Omega_{\eta_{ba}} = \frac{i}{2}(\eps_{AB}\eta_{ba}^A \dd\eta_{ba}^B)\wedge(\eps_{AB}\overline{\eta}_{ba}^A \dd \overline{\eta}_{ba}^B)$.
Then, introducing
\begin{align}
\label{eq:Sprop}
S^{\EPRL} &= \sum_{a<b} \log{\eps(Y^{j^+_{ab},j^-_{ab}}_{j_{ab}}C^{j_{ab}}_{\xi_{ab}},\rho(G_{ab})Y^{j^+_{ab},j^-_{ab}}_{j_{ab}}C^{j_{ab}}_{\eta_{ba}})} \\
S^{\Pi} &= \sum_{a<b} S^{\Pi}_{ab} = \sum_{a<b} \log{\scal{C^{j_{ab}}_{\eta_{ba}}|\Pi_{ba}(\{G_{a'b'}\})C^{j_{ab}}_{\xi_{ba}}} } \\
\Sprop &= S^{\EPRL} + S^{\Pi} ,
\end{align}
we write the amplitude as
\beq
\Aprop\left(\left\lbrace j_{ab}, C^{j_{ab}}_{\xi_{ab}} \right\rbrace\right) =  \int_{\Spin^5} \prod_a \dd G_a \int_{\CP^{10}}  \prod_{a<b}  \dd \mu_{\eta_{ba}} e^{\Sprop} .
\eeq

At this point the reader might expect us to proceed to calculate stationary points of the action. However, we still have to show that stationary phase methods are applicable in this case. The stumbling point is the fact that, while $S^{\EPRL}$ scales linearly with spins $j_{ab}$ (see {\asymptoticE}), $S^{\Pi}$ does not. In what follows, we show that $S^{\Pi}$ is \textit{asymptotically linear} in spins. We employ a strategy similar to the one applied by the author and his collaborators in \cite{evz2016} in the case of the Lorentzian proper vertex amplitude. 

Let $n_{\nu_{ba}} = \beta_{ab}(\{G_{a'b'}\}) \frac{\tr(\sigma^i X^-_{ab}X^+_{ba})}{|\tr(\sigma^i X^-_{ab}X^+_{ba})|}$ and $\nu_{ba}$ be the corresponding unit spinor. Define 
\beq 
\ket{\xi; k,m} = g(\xi)\ket{k,m}.
\eeq
The projector $\Pi_{ba}(\{G_{a'b'}\})$ can be written explicitly
\beq
\Pi_{ba}(\{G_{a'b'}\}) = \Pi_{(0,\infty)} \left(n_{\nu_{ba}} \cdot \hat{L}\right) = \sum^{j_{ab}}_{m>0} \ket{\nu_{ba};j_{ab},m} \bra{\nu_{ba};j_{ab},m}.
\eeq
Then,
\begin{align}
e^{S^{\Pi}_{ab}} &= \scal{C^{j_{ab}}_{\eta_{ba}}|\Pi_{ba}(\{G_{a'b'}\})C^{j_{ab}}_{\xi_{ba}}} \nonumber \\
&= \sum^{j_{ab}}_{m>0} \scal{j_{ab},j_{ab}|g(\eta_{ba})^{-1}g(\nu_{ba})|j_{ab},m}\scal{j_{ab},m| g(\nu_{ba})^{-1} g(\xi_{ba})| j_{ab},j_{ab}}
\end{align}
From this we can use exactly the same argument as in the Lorentzian proper vertex asymptotics paper and obtain (for the details of the argument see \cite{evz2016})\footnote{There is a subtlety here in that these expressions apply to the case of integer $j$. However, similar expressions would pertain in the half-integer case and, in fact, the resulting asymptotics of the vertex amplitude \eqref{eq:propasym} is exactly the same in both cases.}
\begin{align}
\label{eqn:intasym}
\exp (S^\Pi_{ab}) \sim
\left\{ \begin{array}{ll}
(x_{ab}+y_{ab})^{2 \lambda j} & \text{ if }|x_{ab}|>|y_{ab}|\text{ and }|x_{ab}+y_{ab}|^2\ge|4x_{ab}y_{ab}| \\
\frac{(4 x_{ab} y_{ab})^{\lambda j}}{\sqrt{\pi \lambda j}}\frac{x_{ab}}{y_{ab}-x_{ab}} & \text{ if }|x_{ab}|<|y_{ab}|\text{ or }|x_{ab}+y_{ab}|^2<|4x_{ab}y_{ab}|
\end{array} \right.
\end{align}
where $x_{ab} := \scal{\eta_{ba},\nu_{ba}} \scal{\nu_{ba},\xi_{ba}}$ and $y_{ab} := \scal{\eta_{ba},J\nu_{ba}} \scal{J\nu_{ba},\xi_{ba}}$.

Using lemma 4 and theorem 4 in \cite{evz2016} and the analysis of the critical points of the action in {\jonProp} we deduce the asymptotics:
\beq
\label{eq:propasym}
\Aprop\left(\left\lbrace j, C^{j}_{\xi_{ab}} \right\rbrace\right)\Big|_{j\rightarrow j_0} \sim j^{-12} N^{\propdec} \exp\left(i \sum_{a<b} \gamma j \Theta_{ab}\right)\Bigg|_{j\rightarrow j_0}.
\eeq
Here $N^{\propdec}$ is independent of $j$ and $\Theta_{ab}$ are dihedral angles determined by $N_a \cdot N_b = \cos \Theta_{ab}$ with $N_a$, $N_b$ the outward normals to the $a$ and $b$ tetrahedra, respectively. In the case of the regular $4$-simplex, considered in this paper, $\Theta_{ab} \equiv \Theta := \arccos(-\frac{1}{4})$.

Using this asymptotics in \eqref{eq:wpropgauss} and defining $\tilde{z} := z - \gamma \Theta$, we obtain
\beq
\Wprop(z) =  \sum_{j_{ab}} \left(\prod_{a<b} d_{j_{ab}} e^{-t\hbar j_{ab}(j_{ab}+1) - i\tilde{z}j_{ab}}\right)  N_{\{j_{ab}\}}
\eeq
with 
\beq
N_{\{j_{ab}\}} = j_{ab}^{-12} N^{\propdec}\Big|_{j_{ab}\rightarrow j_0} .
\eeq

We can write the prefactor in the explicitly Gaussian form as
\beq
\label{eq:Gaus}
\Wprop(z) \approx  \sum_{j_{ab}} \left(\prod_{a<b} d_{j_{ab}} e^{-t\hbar(j_{ab}-j_0)^2} e^{-\frac{\tilde{z}^2}{4t\hbar}}\right) N_{\{j_{ab}\}}.
\eeq
This is a Gaussian peaking spins at $j_0 = -\frac{i\tilde{z}}{2t\hbar}$ with the spread $\sigma = \frac{1}{\sqrt{2t\hbar}}$. Given that $\Re(j_0) \sim \eta$, in the large volume limit we can therefore approximate the sum by an integral. 
Performing the integration, we get
\beq
\Wprop(z) = \left(\sqrt{\frac{\pi}{t\hbar}} 2j_0 e^{-\frac{\tilde{z}^2}{4t\hbar}}\right)^{10} j_0^{-12} N^{\propdec},
\eeq
and, substituting the definition of $j_0$, we obtain 
\beq
\label{eq:Wres}
\Wprop(z) \approx N \tilde{z}^{-2} e^{\frac{-5\tilde{z}^2}{2t\hbar}} 
\eeq
where $N = -(64\pi)^2 \left(\frac{\pi}{t\hbar}\right)^3 N^{\propdec}$. This is the Hartle-Hawking wavefunction of a closed, homogeneous and isotropic universe.

\section{Classical limit}
\label{sec:classlim}
In this section we will confirm that the Hartle-Hawking state \eqref{eq:Wres} satisfies the Hamiltonian constraint in the classical limit. The Hamiltonian constraint in FRW models is given by \cite{abl2003}:
\beq
C_H = -\frac{3}{8\pi G\gamma^2} c^2 |p|^{\half} \sgn(p) = 0.
\eeq
Rescaling by $|p|^{\frac{3}{2}} \sgn(p)$ we have
\beq
C_H = -\frac{3}{8\pi G\gamma^2} c^2 p^2 = 0.
\eeq
Using \eqref{eq:xieta} and \eqref{eq:zdef},
\beq
\label{eq:ch}
C_H = \frac{3}{128\pi G\gamma^2(\alpha\beta)^2} (z^2 - \bar{z}^2)^2.
\eeq

Let us fix $t$ so that $z$ is a coordinate in the phase space with the symplectic structure 
\beq
\omega = \frac{5i}{t} dz \wedge d\bar{z}.
\eeq
We will discuss below the significance of this choice. Then the Poisson bracket reads $\{z, \bar{z}\} = \frac{it}{5}$. Choosing a holomorphic polarisation for the quantisation, we get the states as holomorphic functions of $z$ and, bearing in mind that $[\hat{z}, \hat{\bar{z}}] = i\hbar\{z, \bar{z}\}$, we define the quantisation of the phase space variables to be
\beq
\hat{z} = z \quad\quad\quad \hat{\bar{z}} = \frac{t\hbar}{5} \frac{d}{dz} - \gamma \Theta
\eeq
These operators satisfy the commutation relations. Furthermore, the adjointness condition $\hat{\bar{z}}^{\dagger} = \hat{z}$ is fulfilled when the Hermitian inner product on the Hilbert space is taken to be
\beq
\langle \psi, \phi \rangle = \int e^{-\frac{5}{t\hbar} (|z|^2 + 2\gamma \Theta \Re{z})} \bar{\psi} \phi \dd^2 z.
\eeq
It is easy to check that the Hartle-Hawking wavefunction $\Wprop(z)$ is normalisable using this inner product.

If we now specify an operator $\hat{H}$ as
\beq
\hat{H} = \frac{3}{128\pi G\gamma^2(\alpha\beta)^2} \left(z^2 - \left(\frac{t\hbar}{5}\frac{d}{dz} - \gamma \Theta \right)^2 +\frac{3t}{5}\hbar + \frac{4t}{5}\hbar \gamma \Theta \tilde{z}^{-1} + \frac{6t^2}{25} \hbar^2 \tilde{z}^{-2}\right)^2,
\eeq
we can note that 
\beq
\hat{H} \Wprop(z) = 0.
\eeq
Considering the limit $\hbar \rightarrow 0$, we see that the classical analogue of $\hat{H}$ is $C_H$. Therefore, we deduce that $\hat{H}$ is a possible quantisation of $C_H$. We can conclude that the Hartle-Hawking wavefunction $\Wprop(z)$ describes a state (with the interpretation of the holomorphic variable $z$ as the coordinate in the phase space defined above) in a quantisation of the FRW model that satisfies the Hamiltonian constraint.

Let us comment on the fact that we had to fix $t$ for this calculation. The choice of $t$ (and, therefore, the spread $\sigma$) amounts to selecting a specific family of coherent states as the boundary states for the amplitude. We can interpret this choice as a restriction imposed by the dynamics of the problem on the allowable set of coherent states. Such dynamic restrictions on coherent states arise in standard quantum mechanics \cite{gk1999}. In what follows we illustrate this with a simple example.

Consider a quantum harmonic oscillator specified by the Hamiltonian
\beq
\label{eq:oscham}
\hat{H} = \frac{\hat{p}^2}{2m} + \frac{m\omega^2 \hat{x}^2}{2}.
\eeq
Define the creation and annihilation operators $\hat{a}_{\kappa}^{\dagger}, \hat{a}_{\kappa}$:
\beq
\hat{a}_{\kappa}^{\dagger} = \sqrt{\frac{\kappa}{2\hbar}} \hat{x} - i \sqrt{\frac{1}{2\hbar\kappa}} \hat{p} \quad\quad \hat{a}_{\kappa} = \sqrt{\frac{\kappa}{2\hbar}} \hat{x} + i \sqrt{\frac{1}{2\hbar\kappa}} \hat{p}
\eeq
The vacuum state $\ket{0_{\kappa}}$ is annihilated by the annihilation operator:
\beq
\label{eq:cohk}
\hat{a}_{\kappa} \ket{0_{\kappa}} = 0.
\eeq
Let $\hat{D}_{\kappa}(\alpha)$ be the unitary displacement operator and define coherent states
\beq
\label{eq:cohpsi}
\ket{\psi_{\kappa}} := \hat{D}_{\kappa}(\alpha) \ket{0_{\kappa}} = e^{\alpha \hat{a}_{\kappa}^{\dagger} - \bar{\alpha}\hat{a}_{\kappa}} \ket{0_{\kappa}}.
\eeq
These coherent states are eigenstates of the annihilation operator,
\beq
\hat{a}_{\kappa} \ket{\psi_{\kappa}} = \alpha \ket{\psi_{\kappa}} ,
\eeq
and saturate the lower bound for the product of uncertainties $(\Delta x)(\Delta p)$ given by the Heisenberg uncertainty principle. These families of coherent states are characterised by the parameter $\kappa$.

It is the dynamics of the harmonic oscillator that fixes the parameter $\kappa$. Using \eqref{eq:cohk} and \eqref{eq:cohpsi} and making $\alpha$ time-dependent, we can define a time-dependent normalised coherent state $\psi_{\kappa}(x,t)$:
\beq
\psi_{\kappa}(x,t) := \left(\frac{\kappa}{\pi\hbar}\right)^{\frac{1}{4}} e^{-\frac{\kappa}{2\hbar}x^2 + \sqrt{\frac{2\kappa}{\hbar}}\alpha(t) x - \alpha(t)\Re{\alpha(t)} + i\phi(t)}
\eeq
where $\phi(t)$ is a phase factor. We now impose that $\psi_{\kappa}(t)$ satisfy the Schr\"odinger equation:
\beq
\label{eq:schreq}
\hat{H} \ket{\psi_{\kappa}(t)} = i\hbar\frac{d}{dt}\ket{\psi_{\kappa}(t)}.
\eeq
This yields for the harmonic oscillator characterised by the Hamiltonian \eqref{eq:oscham}:
\beq
\left(-\frac{\hbar^2}{2m}\frac{d^2}{dx^2} + \frac{m\omega^2 x^2}{2}\right) \psi_{\kappa}(x,t) = i\hbar \frac{d}{dt}\psi_{\kappa}(x,t).
\eeq
Solving this condition gives $\kappa = m\omega$. Hence, the dynamics of the problem restricts the family of the coherent states to the canonical coherent states associated with the quantum harmonic oscillator.
 
Similarly to \eqref{eq:schreq}, in the case of quantum cosmology the condition being imposed reads:
\beq
\hat{H} \ket{\Psi_{t}} = 0.
\eeq
This condition, as we have seen, likewise selects a specific family of heat-kernel coherent states by fixing $t$ and, therefore, the spread $\sigma$.

\section{Long-range correlations}
\label{sec:corr}
As we have seen above, the dynamics selects a family of heat-kernel coherent states by imposing a condition on the heat-kernel time $t$ which is equivalent to constraining the width $\sigma$ of the Gaussian in \eqref{eq:Gaus}. It is natural to investigate whether there is any similar constraint on the long-range correlations. However, the most widely considered coherent states use the Laplace-Beltrami operator as a complexifier \cite{thiemann2001, thiemann2006a} which does not generate off-diagonal elements in the covariance matrix between spins associated to different links of the boundary graph. In fact, there is no complexifier in literature that readily yields complexifier coherent states with long-range correlations (it has been recently proposed to use "squeezed" coherent states to remedy this issue \cite{bghy2016}; however, these are not complexifier coherent states).

We can investigate the long-range correlations in the present set-up by replacing the pre-factor in \eqref{eq:wpropgauss} as follows:
\beq
\label{eq:wpropcor} 
\Wcor(z) =  \sum_{j_{ab}} \left(\prod_{a<b} d_{j_{ab}} e^{ -\hbar \sum_{c<d} P^{(ab)(cd)} j_{ab}j_{cd} - izj_{ab}}\right) \Aprop\left(\left\lbrace j_{ab}, C^{j_{ab}}_{\xi_{ab}} \right\rbrace\right) .
\eeq

Here we introduced the covariance matrix $P$ which has the symmetries of the regular 4-simplex and can therefore be written as
\beq
P = \sum^3_{i=1} \rho_i P_i,
\eeq
where $\rho_i$ are three real numbers and the matrices $P_1, P_2, P_3$ have the following form:
\begin{itemize}
\item $P_1^{(ab)(cd)} = 1$ if $(ab)=(cd)$, $0$ otherwise
\item $P_2^{(ab)(cd)} = 1$ if $a=c$ and $b \neq d$ (and permutations thereof), $0$ otherwise
\item $P_3^{(ab)(cd)} = 1$ if $(ab),(cd)$ are disjoint, $0$ otherwise
\end{itemize}

It is easy to see that this ansatz ensures that the coherent state is peaked on the intrinsic and  extrinsic geometry of the $3$-manifold $M$. The heat-kernel time $t$ has been absorbed into the scaling of $P$. Furthermore, this ansatz is analogous to Rovelli's original proposal \cite{rovelli2006} used in the graviton propagator calculations {\propcit}.

We can rewrite \eqref{eq:wpropcor} in the following form
\beq
\label{eq:wpropcorsq} 
\Wcor(z) =  \sum_{j_{ab}} \left(\prod_{a<b} d_{j_{ab}} e^{ \sum_{c<d} (-\hbar P^{(ab)(cd)} j_{ab}j_{cd} - iz' P^{(ab)(cd)} j_{cd})}\right) \Aprop\left(\left\lbrace j_{ab}, C^{j_{ab}}_{\xi_{ab}} \right\rbrace\right),
\eeq
where we introduced $z' = \frac{z}{\rho} = \frac{\xi}{\rho} + i \frac{\eta}{\rho}$ with $\rho =\rho_1+6\rho_2+3\rho_3$. This allows us to complete the square, perform the calculation as in \autoref{ssec:calc} and obtain (in place of \eqref{eq:Gaus})
\beq
\Wcor(z) \approx  \sum_{j_{ab}} \left(\prod_{a<b} d_{j_{ab}} e^{-\sum_{c<d} \hbar P^{(ab)(cd)}(j_{ab}-j_0)(j_{cd}-j_0)}\right) e^{-\frac{5\tilde{z}^2}{2\rho\hbar}} N_{\{j_{ab}\}},
\eeq
with $j_0 = -\frac{i\tilde{z}}{2\rho\hbar}$. Approximating the sum over spins as a Gaussian integral and performing the integration we get the result
\beq
\Wcor(z) \approx N' \tilde{z}^{-2} e^{\frac{-5\tilde{z}^2}{2\rho\hbar}}
\eeq
where $N' = -(\frac{16}{\hbar})^3 \frac{\pi^5 \rho^2 N^{\propdec}}{\sqrt{\det{P}}}$.

The resulting amplitude has the exact same dependence on $z$ as in \eqref{eq:Wres} with $\rho$ playing the role of $t$. Therefore, the analysis of \autoref{sec:classlim} carries through, and we have
\beq
\hat{H}' \Wcor(z) = 0
\eeq
for an operator $\hat{H}'$,
\beq
\hat{H}' = \frac{3}{128\pi G\gamma^2(\alpha\beta)^2} \left(z^2 - \left(\frac{\hbar\rho}{5}\frac{d}{dz} - \gamma \Theta \right)^2 +\frac{3\rho}{5}\hbar + \frac{4\rho}{5}\hbar  \gamma \Theta \tilde{z}^{-1} + \frac{6\rho^2}{25} \hbar^2\tilde{z}^{-2}\right)^2.
\eeq
As before, it is clear that $\hat{H}'$ is a possible quantisation of $C_H$ and the interpretation of the wavefunction $\Wcor(z)$ is the same as $\Wprop(z)$ above. However, now the wavefunction $\Wcor(z)$ encodes long-range correlations. We obtain that these correlations are restricted by the dynamics to be parametrized by a two-dimensional subspace in $\R^3$ where $\rho_1+6\rho_2+3\rho_3 = -\frac{80\pi G\gamma\alpha\beta}{3}$.

\section{Conclusion}

In this work we investigated spinfoam cosmology using the recently introduced proper vertex amplitude. We evaluated the Hartle-Hawking wavefunction as a spinfoam transition amplitude from a zero three-geometry to a finite three-geometry. To perform the calculation we introduced a fixed graph thereby truncating the boundary Hilbert space of the theory. On this graph we considered coherent boundary states peaked on the classical spatial geometry and extrinsic curvature of the FRW model. The spinfoam expansion was approximated by a single proper vertex. We analysed the asymptotics of the proper vertex for large spins and obtained the associated transition amplitude in the large volume limit. We note here that this transition amplitude would have the same functional form if we were to replace the proper vertex with the EPRL vertex because in the stationary phase analysis of the EPRL vertex the form of the coherent state would select the exact same orientation as that present in the asymptotics of the proper vertex and suppress the other orientation.

The amplitude has been shown to satisfy an operator constraint. This operator constraint can be viewed as a quantisation of the classical Hamiltonian constraint arising in LQC. Note, however, that the dynamics is rather trivial: the spacetime is flat, which is the unique non-degenerate solution of Einstein's equations in the absence of matter and cosmological constant. We found that the dynamics imposes a restriction on the relevant family of coherent states. This is not surprising, because such restrictions arise in standard quantum mechanics. We demonstrated a similar coherent state selection on the example of a quantum harmonic oscillator. We also considered a boundary coherent state with long-range correlations and obtained that the dynamics similarly restricts the parameter space for these correlations.

There are multiple avenues for further investigations: one could include matter or cosmological constant (see \cite{bkrv2011, vidotto2011} for previous work in the EPRL model). One could consider larger graphs and higher orders in the vertex expansion to check the validity of approximations. Another task would be to apply the Lorentzian proper vertex to spinfoam cosmology. Since by construction the boundary data is that of a Euclidean $4$-simplex, only critical points in the degenerate sector are selected {\asymptoticL}. Therefore, the corresponding contributions are expected to be suppressed in the asymptotics of the proper vertex \cite{evz2016}, and the calculation will have to include the next order in the vertex expansion.

\begin{acknowledgments}
The author thanks Jonathan Engle for useful discussions and helpful comments on the draft of this paper. The author wishes to also express gratitude to Carlo Rovelli and Francesca Vidotto for stimulating conversations during the early stages of this project and to Eugenio Bianchi for his suggestion to include long-range correlations in the analysis. This work was partially supported by the National Science Foundation through grants PHY-1205968 and PHY-1505490.
\end{acknowledgments}

\bibliographystyle{apsrev4-1}
\bibliography{ilyabib}

\end{document}